\documentclass[preprint2]{aastex63}
\usepackage{bm}

\newcommand{\msun}{\,\mathrm{M}_\odot}
\usepackage{subfigure}

\received{October 1, 2020}
\revised{February 23, 2021}
\accepted{February 23, 2021}
\submitjournal{ApJL}

\shorttitle{Mass and density of asteroid (16) Psyche}
\shortauthors{Siltala \& Granvik}
\graphicspath{{./}{figures/}}
\begin{document}

\title{Mass and density of asteroid (16) Psyche}

\correspondingauthor{Lauri Siltala}
\email{lauri.siltala@helsinki.fi}

\author[0000-0002-6938-794X]{Lauri Siltala}
\affiliation{Department of Physics, P.O. Box 64, FI-00014 University of Helsinki, Finland}
\author[0000-0002-5624-1888]{Mikael Granvik}
\affiliation{Department of Physics, P.O. Box 64, FI-00014 University of Helsinki, Finland}
\affiliation{Asteroid Engineering Laboratory, Onboard Space Systems, Luleå University of Technology, Box 848, S-98128 Kiruna, Sweden}

%% Mark off the abstract in the ``abstract'' environment. 
\begin{abstract}

We apply our novel Markov-chain Monte Carlo (MCMC)-based algorithm for asteroid
  mass estimation to asteroid (16) Psyche, the target of NASA's eponymous Psyche
  mission, based on close encounters with 10 different asteroids, and obtain
  a mass of $(1.117 \pm 0.039) \times 10^{-11} \msun$. We ensure that our
  method works as expected by applying it to asteroids (1) Ceres and (4) Vesta,
  and find that the results are in agreement with the very accurate mass
  estimates for these bodies obtained by the Dawn mission. We then combine our
  mass estimate for Psyche with the most recent volume estimate to compute the
  corresponding bulk density as $(3.88 \pm 0.25)$ g/cm$^3$. The estimated bulk
  density rules out the possibility of Psyche being an exposed, solid iron core
  of a protoplanet, but is fully consistent with the recent hypothesis that
  ferrovolcanism would have occurred on Psyche. 

\end{abstract}

%% Keywords should appear after the \end{abstract} command. 
%% See the online documentation for the full list of available subject
%% keywords and the rules for their use.
%% From the front matter, we move on to the body of the paper.
%% Sections are demarcated by \section and \subsection, respectively.
%% Observe the use of the LaTeX \label
%% command after the \subsection to give a symbolic KEY to the
%% subsection for cross-referencing in a \ref command.
%% You can use LaTeX's \ref and \label commands to keep track of
%% cross-references to sections, equations, tables, and figures.
%% That way, if you change the order of any elements, LaTeX will
%% automatically renumber them.
%%
%% We recommend that authors also use the natbib \citep
%% and \citet commands to identify citations.  The citations are
%% tied to the reference list via symbolic KEYs. The KEY corresponds
%% to the KEY in the \bibitem in the reference list below. 

\section{Introduction} 
Asteroid (16) Psyche is currently of great interest to the planetary science community as it is the target of NASA's eponymous Psyche mission currently scheduled to be launched in 2022 \citep{Elk14}. This interest stems from the fact that the asteroid is currently believed to be a metallic asteroid and potentially the exposed core of a protoplanet due to both its relatively high bulk density as well as surface properties based on spectroscopic and radar observations \citep{She17}. There have, however, been concerns that Psyche's relatively high bulk density of approximately 4 g/cm$^3$ is still too low to be consistent with iron meteorites and that the asteroid may instead have a stony-iron composition and could thus be a parent body for mesosiderites \citep{Vii18}. Ferrovolcanic activity has recently been suggested as an alternative mechanism that would explain the observational data on Psyche \citep{Joh20}. Ferrovolcanism would cause Psyche's surface to consist of stony mantle surrounded by a metallic surface layer resulting from past eruptions of molten iron. This theory is consistent with both a relatively low bulk density and a metal-rich surface composition.

We have developed a Markov-chain Monte Carlo (MCMC)-based algorithm for asteroid mass estimation based on asteroid-asteroid close encounters. In essence, the gravitational perturbations of a massive asteroid on the orbit of another asteroid with negligible mass (hereafter referred to as the test asteroid) are modeled by fitting orbits for both objects and the mass for the massive asteroid so that they accurately reproduce the typically extensive number of astrometric observations available. In \citet{Sil20}, we applied our algorithm to Psyche (among several other asteroids) and obtained maximum-likelihood (ML) masses approximately half of the average literature value of $1.37 \times 10^{-11} \msun$ directly leading to a significantly lower bulk density for Psyche than previously reported \citep{Car12}. However, the aforementioned average literature value remained within our $3\sigma$ limits and could thus not be entirely ruled out. We also predicted that accurate astrometry obtained during the summer of 2019 of one of the test asteroids, (151878) 2003 PZ4, could significantly reduce the uncertainty in our mass estimate for Psyche.

Here we use our MCMC method to re-estimate the mass of (16) Psyche by simultaneously including 10 test asteroids as opposed to our previous work where we performed two separate runs where each run simultaneously modeled two separate test asteroids. The use of 10 test asteroids dramatically increases the amount of observational data included and is expected to accordingly lead to a reduced uncertainty for the mass estimate. To test the aforementioned prediction, we have also obtained astrometry of (151878) 2003 PZ4 in July and August of 2019 with the 2.56-meter Nordic Optical Telescope and include this new data in the analysis.

\section{Methods} 
We will describe our algorithm only briefly, as it is already documented in greater detail in our previous work \citep{Sil20}.
We estimate an asteroid's (hereafter called the perturber) mass by modeling its gravitational perturbation on another asteroid (hereafter called test asteroid) during a close encounter between the two. The approach can be seen as an extension of the orbit determination problem where we simultaneously fit orbits for both the perturber and the test asteroids as well as the mass of the perturber requiring that the solution allow us to reproduce the astrometry available for each object. Our approach is based on the Robust Adaptive Metropolis algorithm \citep{Vih12} which can be seen as a Metropolis-Hastings MCMC algorithm with the addition that the proposal distribution is adapted after each proposal with the intent of optimizing the MCMC acceptance rate.

The only update to the algorithm presented in \citet{Sil20} is a slightly different acceptance criterion; where previously we computed the acceptance criterion with a single posterior probability density value based on the sum of the $\chi^2$ values for all targets, we now consider the $\chi^2$ values separately for each target and compute their product to obtain the final acceptance probability:  % p(\bm{P}') \propto \exp(-\frac{1}{2} \chi^2(\bm{P}')) \,.
\begin{equation}
      a_i = \prod_j \frac{p_j(\bm{P}')}{p_j(\bm{P}_i)} = \prod_j \exp(-\frac{1}{2} \left(\chi^2(\bm{P_j}') - \chi^2(\bm{P_j})\right))
\end{equation}
where $j$ represents each individual asteroid considered. Such an approach ensures that accepted proposals must fit each individual asteroid well as opposed to the  criterion used by \citet{Sil20}, which only required that the overall fit be acceptable.
In addition, this means that each asteroid has the same weight in the acceptance criterion while with our previous approach the total $\chi^2$ value and, by extension, acceptance criterion, is dominated by asteroids with more observations as such have a greater impact on the total value. We use a uniform prior of unity \citep[see][and discussion therein]{Sil20}.

\section{Data} 

Our data treatment also follows the same approach as in our previous work; we use data obtained from the Minor Planet Center, correct for star catalogue biases \citep{Far15}, apply an observational error model to weight the data properly \citep{Bae17}, and multiply the uncertainties for $N$ same-night observations by a factor of $\sqrt{N}$ such that an individual observation's weight is $1/(N\sigma^2 )$ where $\sigma$ is the corresponding astrometric uncertainty based on the error model.
In addition to Psyche itself we use the following 10 numbered test asteroids: (1052), (1082), (6442), (13206), (17799), (20837), (39054), (91495), (151878) and (211012).
We have already previously studied four of these test asteroids, namely (17799), (20837), (91495), and (151878) \citep{Sil20} whereas the other test asteroids have been identified and/or used by \citet{Fie03,Bae17,Gal02}.  Test asteroid (1054) is particularly noteworthy as it has six close encounters with Psyche during the observational timespan \citep{Fie03}.
We use all of the astrometry available through the MPC taken between the start of 1980 and Oct 2020 for our objects for which the star catalogue debiasing can be applied (i.e., the observations that include information on the star catalogue in use). We make an exception for the asteroid (13206), for which we use all available data, as its encounter with Psyche took place in 1974 \citep{Gal02}. Rejecting astrometry that cannot be debiased is a conscious decision on our part with the intent of avoiding potential issues rising from combining debiased and biased observations. Correspondingly, we have taken care to only select test asteroids for which there exist enough debiased data both before and after the asteroid's close encounter with Psyche.

\citet{Sil20} computed future ephemerides for each accepted proposal in an MCMC chain for Psyche with the test asteroids (91495) and (151878) and used these to show that the mass of Psyche would have a particularly large impact on the sky coordinates of the asteroid (151878) during the summer of 2019. In order to test the prediction, we also obtained astrometry for this asteroid with the Nordic Optical Telescope on La Palma to be used in the mass estimation for Psyche. Five images with an exposure time of 60 seconds each were taken on the 24th and the 29th of July for a total of ten images, nine of which were taken with the R Bessel filter whereas the first image was taken with the V filter due to observer error but was nevertheless useful for astrometry. Both nights had a single observation each where the target overlapped with a background star, rendering the observation useless. Thus a total of eight of these observations were used. The data was reduced using IRAF for bias and flat-field corrections while astrometric processing was done using the Astrometry.net software \citep{Lan10}.

\begin{figure}
\begin{center}
    \includegraphics[width=1.0\columnwidth]{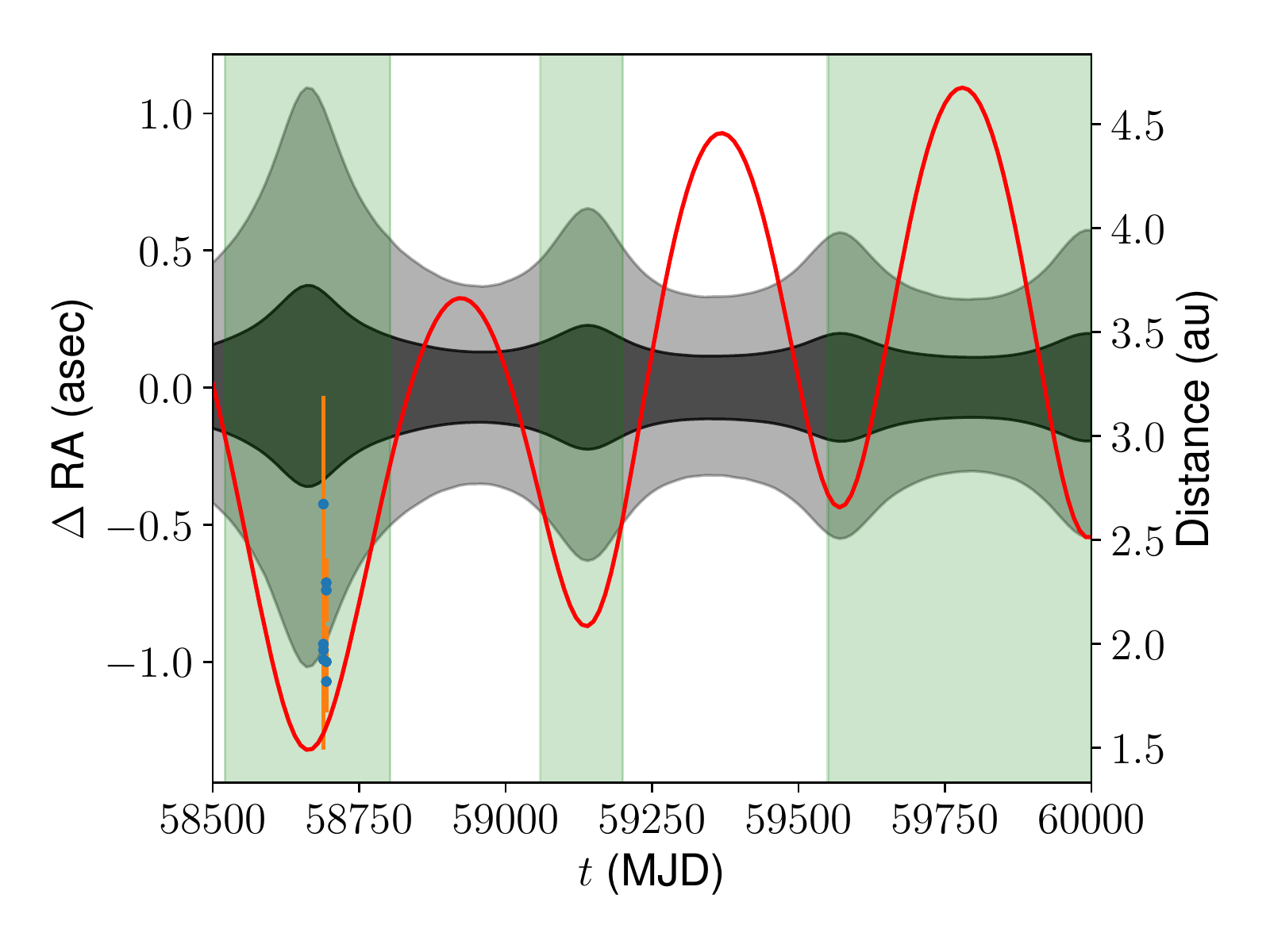}
    \caption{Ephemeris prediction for asteroid (151878) 2003 PZ$_4$ up to MJD 60000 in terms of RA relative to the best-fit value \citep{Sil20}. The $1\sigma$ and $3\sigma$ credible intervals, that account for astrometric uncertainties and the uncertainty of Psyche's mass, are shown in darker and lighter gray, respectively. The red line shows the asteroid's topocentric distance as a function of time. The green color represents times when the asteroid is observable assuming a topocentric observer by requiring that the solar elongation is greater than 60 degrees and the apparent V magnitude is less than 21.
    The blue data points represent the NOT astrometry with orange error bars. In calendar dates the timespan ranges from Jan 17, 2019 to Feb 25, 2023.}
    \label{151878_ra_prediction}
  \end{center}
\end{figure}

Figure~\ref{151878_ra_prediction} shows these observations and their astrometric uncertainties (computed based the uncertainty of the plate solutions combined with the PSF of the object) overlaid on the prediction by \citet{Sil20} in terms of Right Ascension (RA). It is clear that each observation has a smaller RA than the one corresponding to the ML value, which was predicted to translate to a lower-than-nominal mass for Psyche.

To test our mass-estimation method we compute mass estimates for (1) Ceres with the test asteroids (5303) and (46938) and (4) Vesta with the test asteroids (8331) and (125655). Each of these encounters has been previously studied by, e.g., \citet{Bae17}. Both Ceres and Vesta have very accurate mass estimates from the Dawn mission \citep{Rus16, Rus12} and the remaining uncertainties can be considered negligible for the purposes of testing our method.

\section{Results and discussion}

\begin{figure*}
    \centering
\begin{subfigure}
\centering
    \includegraphics[width=1.0\columnwidth]{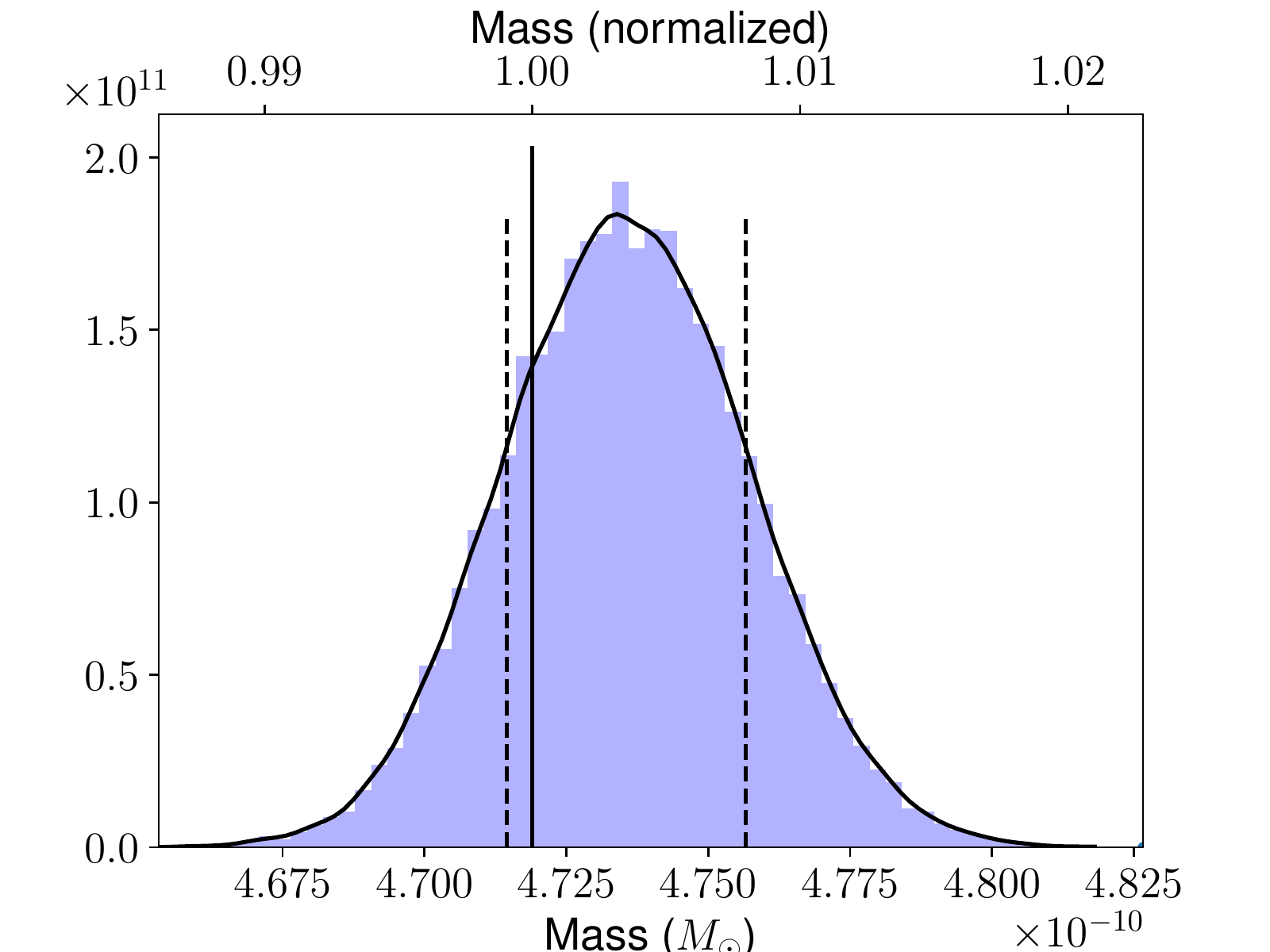}
\end{subfigure}
\begin{subfigure}
\centering
    \includegraphics[width=1.0\columnwidth]{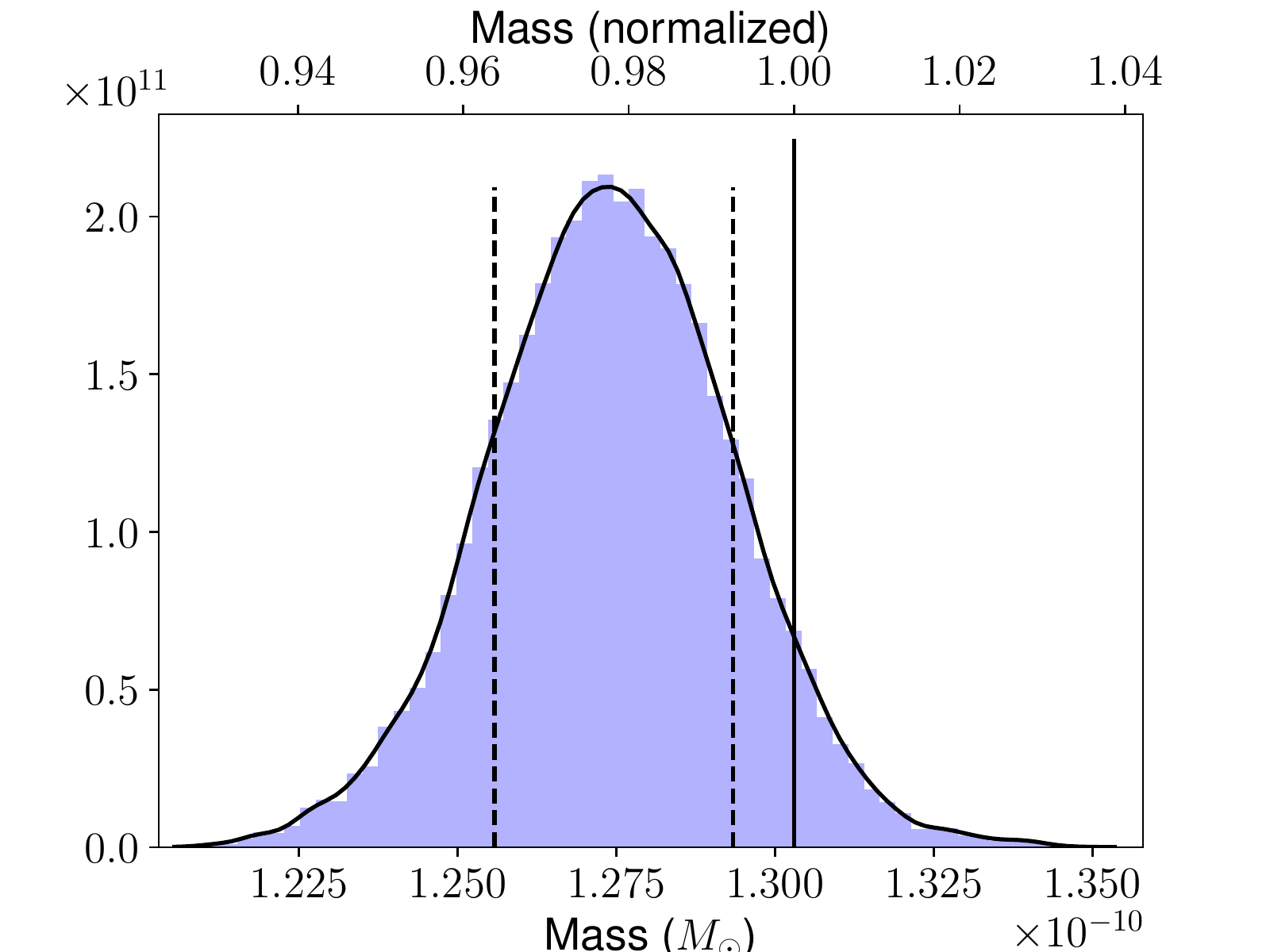}
\end{subfigure}
\caption{Probability distribution for the masses of (1) Ceres (left) and (4) Vesta (right). The upper x axes show the mass relative to Dawn's estimates of $4.719 \times 10^{-10} \msun$ \citep{Rus16} and $1.303 \times 10^{-10} \msun$ \citep{Rus12}, respectively, which have low enough uncertainties that they can be ignored here. The dashed vertical lines represent our $1\sigma$ limits whereas the black curves represent kernel-density estimates fitted to the normalized histograms. The solid black lines correspond to the mass from Dawn.}
\label{CeresandVesta}
\end{figure*}

Let us first test our method by applying it to two asteroids with very accurately known masses. 
The Dawn estimates for the masses of Ceres and Vesta are $(4.7192 \pm 0.0005) \times 10^{-10 }\msun$ and $(1.302891 \pm 0.000005) \times 10^{-10}\msun$, respectively \citep{Rus16, Rus12}. With two test asteroids in each case, we obtain a mass of $(4.73 \pm 0.02) \times 10^{-10} \msun$ for Ceres and $(1.27 \pm 0.02) \times 10^{-10} \msun$ for Vesta (Fig.~\ref{CeresandVesta}). For Ceres the Dawn results are within our $1\sigma$ limits while for Vesta Dawn's mass estimate falls within our $2\sigma$ limits. We get very close to the expected Dawn values in absolute terms and statistically the results are also expected, and thus the test results suggest that our method produces reliable results.

\begin{figure}
  \begin{center} 
    \includegraphics[width=1.0\columnwidth]{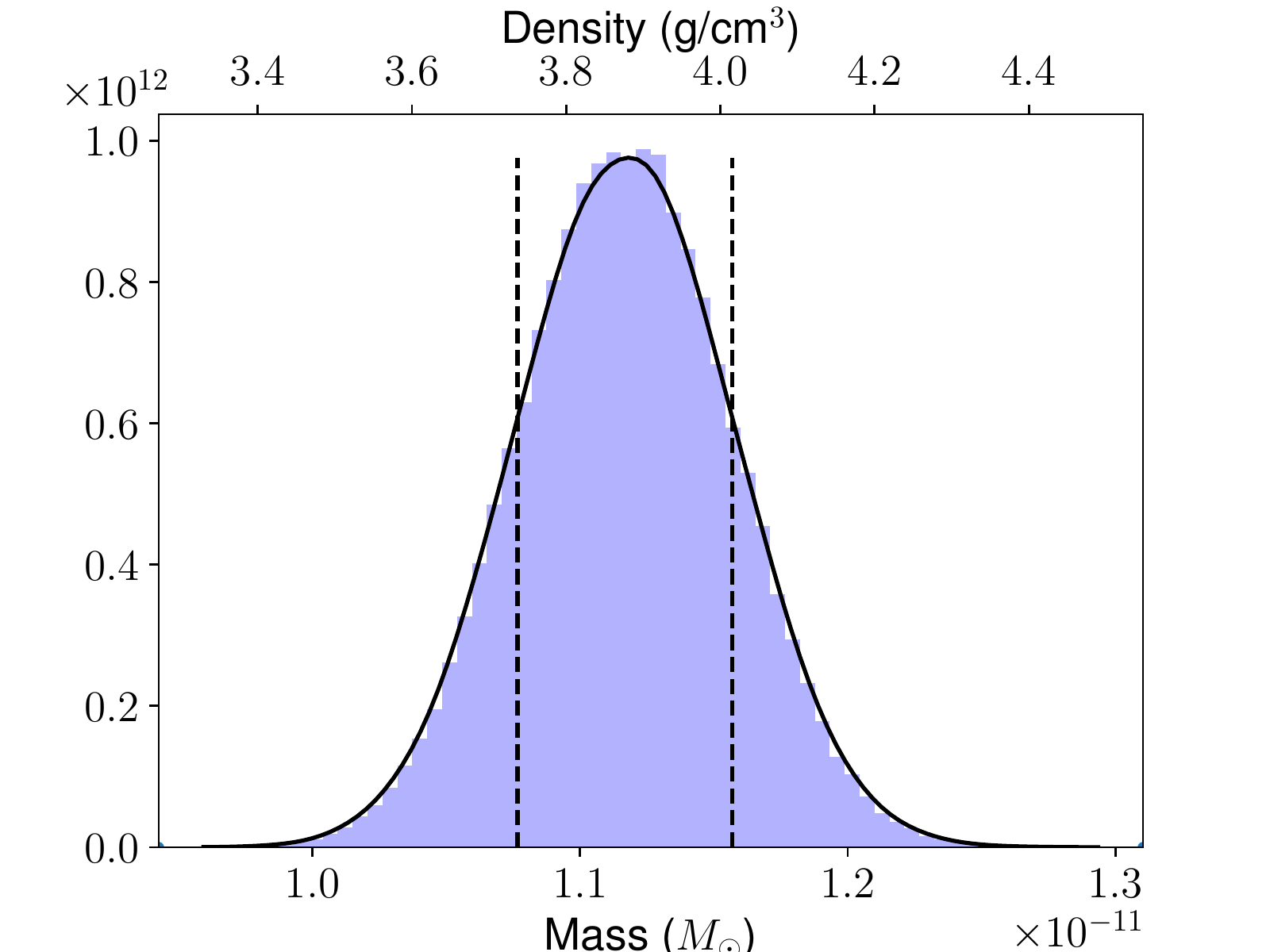}
    \caption{Probability distribution for the mass of Psyche. The upper x-axis shows the bulk density that corresponds to the mass on the lower x axis assuming a volume-equivalent diameter of 222 km. We note that the bulk density does not take the diameter's uncertainty into account. The dashed vertical lines represent our $1\sigma$ limits whereas the black curve represents a kernel density estimate fitted to the normalized histogram.}
    \label{psyche_distro}
  \end{center}
\end{figure}

Now let us turn our attention to (16) Psyche. Figure~\ref{psyche_distro} shows the probability distribution for the mass of Psyche. %in the form of a histogram of each accepted MCMC proposal weighted with the number of times each proposal was accepted.  
From visual inspection it is apparent that the distribution is quite symmetric and Gaussian. We fitted a kernel density estimate on the results, based on which we obtain a mass of $(1.117 \pm 0.039 \times 10^{-11}) \msun$ corresponding to a gravitational parameter (GM) of $(1.482 \pm 0.052)$ km$^3$/s$^2$. In comparison, \citet{Bae17} obtained a mass of $(1.15 \pm 0.035 \times 10^{-11}) \msun$
for (16) Psyche based on the test asteroids (13206), (211012) and (39054), all of which were also included in this work. Clearly, despite the significantly larger number of data used in this study, our uncertainties remain slightly wider than those of \citet{Bae17}. This can be explained by the $\sqrt{N}$ factor used in our weighting of the data which directly leads to wider uncertainties as also demonstrated in \citet{Sil20}.

\begin{figure*}[hbtp]
  \begin{center} 
  \hspace*{-1cm}%
    \includegraphics[width=2.5\columnwidth]{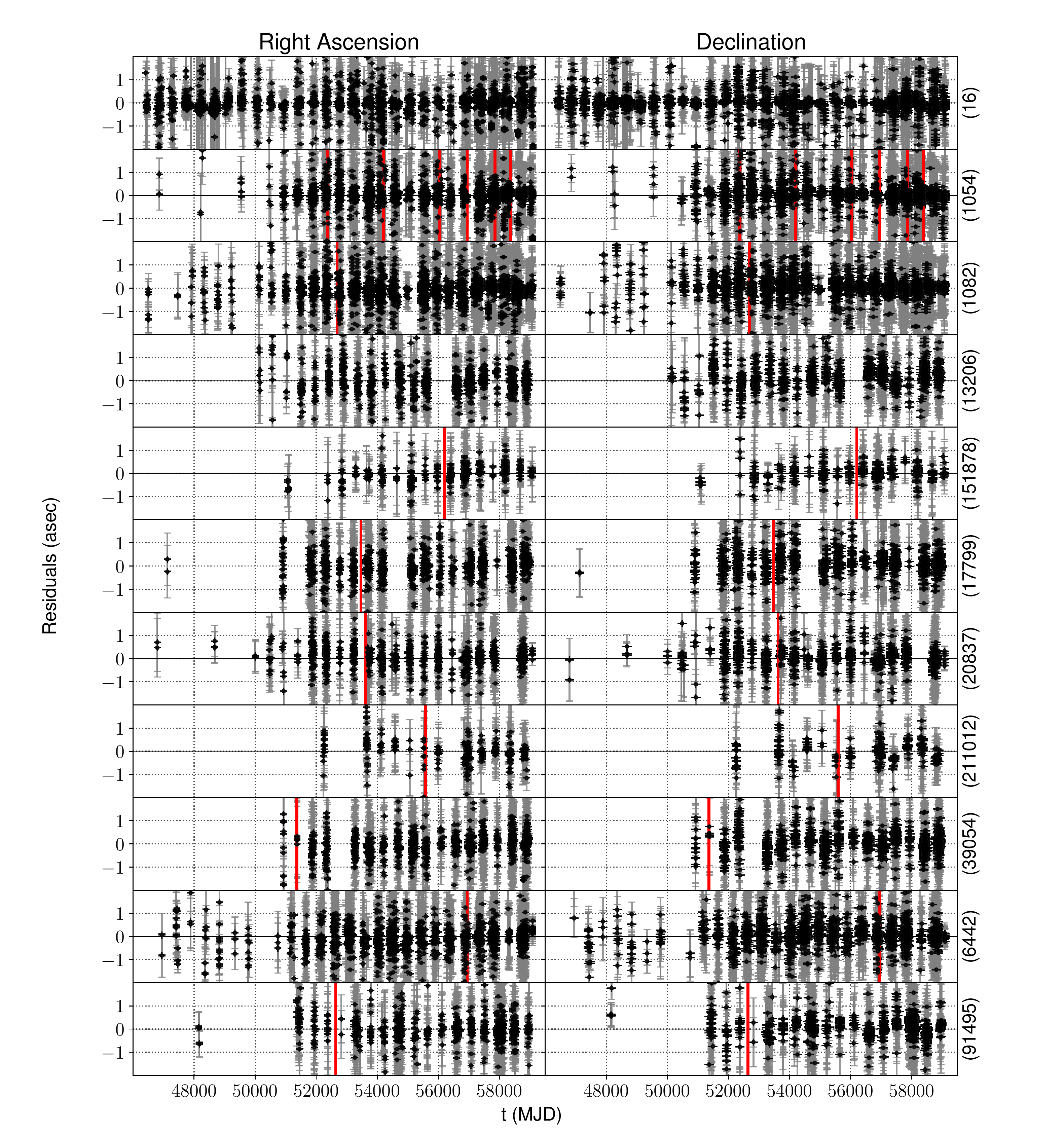}
    \caption{Residuals for (16) Psyche and each test asteroid corresponding to the ML solution. The grey error bars represent the noise assumption for each observation whereas the black error bars represent the $1\sigma$ scatter of the residuals for all accepted proposals. The solid vertical red lines represent the epochs of the close encounters each individual test asteroid had with Psyche. In calendar dates the timespan ranges from Jan 01, 1980 to Oct 13, 2021.}
    \label{resids}
  \end{center}
\end{figure*}

\begin{figure}
  \begin{center} 
    \includegraphics[width=1.0\columnwidth]{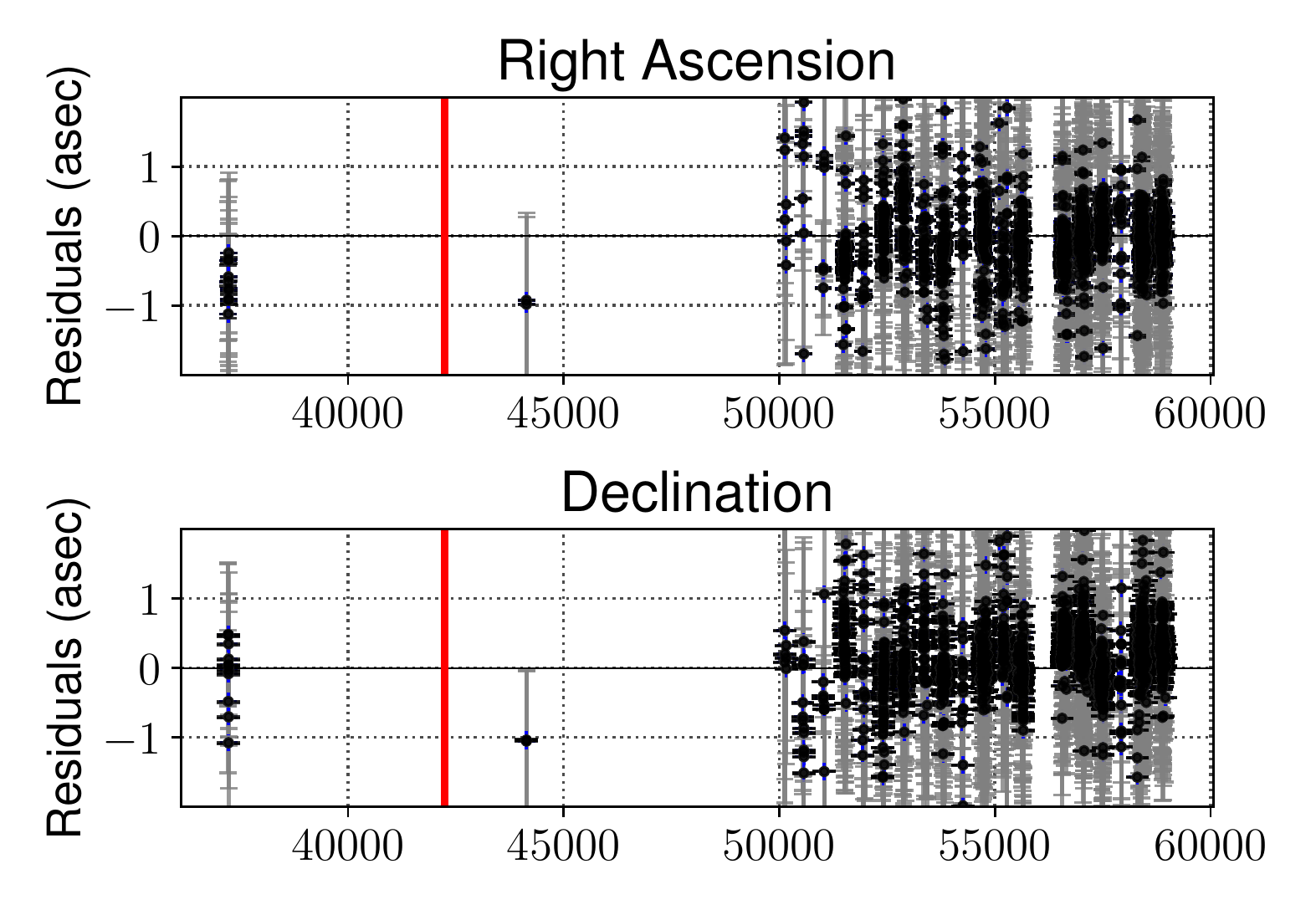}
    \caption{Residuals for (13206) Baer corresponding to the ML solution. The grey error bars represent the noise assumption for each observation whereas the black error bars represent the $1\sigma$ scatter of the residuals for all accepted proposals. The solid vertical red lines represent the epoch of the asteroid's close encounter with Psyche. In calendar dates the timespan ranges from
    Oct 25, 1957 to Apr 29, 2023. }
    \label{13206_resids}
  \end{center}
\end{figure}

To gauge the goodness of fit directly, Fig.~\ref{resids} includes the residuals of each asteroid corresponding to the ML solution in addition to the epochs of each test asteroid's close encounter with Psyche, 
with the exception of (13206) due to the relevant encounter being far in the past. For that particular asteroid, residuals across the entire timespan are included separately in Fig.~\ref{13206_resids}.
There are no clear systematic effects seen in the residuals. The prediction by \citet{Sil20} for the correlation between RA residuals for (151878) and Psyche mass suggest that, based on the NOT astrometry, the mass of Psyche should range from zero to $10^{-11} \msun$, which is in agreement with the above value. 

Recently, \citet{Fer20} reported a volume-equivalent diameter of $(222 \pm 4)$ km for Psyche. Based on this diameter, our mass estimate corresponds to a bulk density of $(3.88 \pm 0.25)$ g/cm$^3$, taking into account both the uncertainties of the mass and the volume-equivalent diameter. In comparison, \citet{Vii18} recently reported a bulk density of $(3.99 \pm 0.26)$ g/cm$^3$ for this object whereas \citet{Sil20} recently obtained bulk densities of $(2.68 \pm 1.21)$ g/cm$^3$ and $(2.54 \pm 0.98)$ g/cm$^3$ based on two independent mass estimates. Hence, our new result are within $2\sigma$ of the \citet{Sil20} values while also agreeing with the \citet{Vii18} value within $1\sigma$. It is clear that the inclusion of additional test asteroids and, by extension, additional observational data, has significantly reduced the uncertainties, which was expected.
Overall, it appears that Psyche's bulk density may indeed
be slightly lower than previously believed but not quite as low as the results by \citet{Sil20} suggested.

According to \citet{Vii18}, iron meteorites have a bulk density of about 7.8 g/cm$^3$ which our bulk density estimate (and also previous estimates) strongly disagrees with. It thus appears difficult for Psyche's composition to match such meteorites unless it is highly porous. %which, in our opinion, appears highly unlikely yet cannot conclusively be ruled out here. 
The same authors reported a bulk density of about 4.25 g/cm$^3$ for the stony-iron mesosiderites and noted that Psyche has a similar bulk density. Based on our results that, too, appears statistically unlikely yet cannot be ruled out as such a density remains within our $3\sigma$ limits as seen in Fig.~\ref{psyche_distro}. On the other hand, our results are fully consistent with the ferrovolcanism model proposed by \citet{Joh20} which, depending on Psyche's exact composition and interior structure, permits densities even below 3 g/cm$^3$.


\section{Conclusions}
We compute a mass estimate of $(1.117 \pm 0.039) \times 10^{-11} \msun$ for asteroid (16) Psyche which corresponds to a bulk density of $(3.88 \pm 0.25) $ g/cm$^3$. This is lower than reported by most recent studies yet not quite as low as in \citet{Sil20}. We find that the bulk density is in line with the recent ferrovolcanism hypothesis by \citet{Joh20}. We expect that astrometry (from, e.g., the Gaia mission) will provide further constraints on the mass and bulk density in the future. 
In addition, we have successfully tested our algorithm by obtaining masses of of $(4.73 \pm 0.02) \times 10^{-10} \msun$ and $(1.27 \pm 0.02) \times 10^{-10} \msun$ for Ceres and Vesta, respectively, that are in agreement with the accurate estimates produced by the Dawn mission.

\acknowledgments
This work was supported by grants \#299543, \#307157, and \#328654 from the Academy of Finland. This research has made use of NASA's Astrophysics Data System, and data and/or services provided by the International Astronomical Union's Minor Planet Center. The computations for this work were mainly performed with the Kale cluster of the Finnish Grid and Cloud infrastructure (persistent identifier: urn:nbn:fi:research-infras-2016072533). The data presented here were obtained in part with ALFOSC, which is provided by the Instituto de Astrofisica de Andalucia (IAA) under a joint agreement with the University of Copenhagen and NOTSA. Partially based on observations made with the Nordic Optical Telescope, operated by the Nordic Optical Telescope Scientific Association at the Observatorio del Roque de los Muchachos, La Palma, Spain, of the Instituto de Astrofisica de Canarias.
%% To help institutions obtain information on the effectiveness of their 
%% telescopes the AAS Journals has created a group of keywords for telescope 
%% facilities.
%
%% Following the acknowledgments section, use the following syntax and the
%% \facility{} or \facilities{} macros to list the keywords of facilities used 
%% in the research for the paper.  Each keyword is check against the master 
%% list during copy editing.  Individual instruments can be provided in 
%% parentheses, after the keyword, but they are not verified.

\vspace{5mm}
\facilities{NOT(ALFOSC):2.56m}

%% Similar to \facility{}, there is the optional \software command to allow 
%% authors a place to specify which programs were used during the creation of 
%% the manuscript. Authors should list each code and include either a
%% citation or url to the code inside ()s when available.

\software{OpenOrb \citep{Gra09},  
          Astrometry.net \citep{Lan10},
          IRAF (http://iraf.noao.edu/)
          }

%% For this sample we use BibTeX plus aasjournals.bst to generate the
%% the bibliography. The sample63.bib file was populated from ADS. To
%% get the citations to show in the compiled file do the following:
%%
%% pdflatex sample63.tex
%% bibtext sample63
%% pdflatex sample63.tex
%% pdflatex sample63.tex

\bibliography{refs.bib}{}

\begin{thebibliography}{}
\expandafter\ifx\csname natexlab\endcsname\relax\def\natexlab#1{#1}\fi
\providecommand{\url}[1]{\href{#1}{#1}}
\providecommand{\dodoi}[1]{doi:~\href{http://doi.org/#1}{\nolinkurl{#1}}}
\providecommand{\doeprint}[1]{\href{http://ascl.net/#1}{\nolinkurl{http://ascl.net/#1}}}
\providecommand{\doarXiv}[1]{\href{https://arxiv.org/abs/#1}{\nolinkurl{https://arxiv.org/abs/#1}}}

\bibitem[{{Baer} \& {Chesley}(2017)}]{Bae17}
{Baer}, J., \& {Chesley}, S.~R. 2017, \aj, 154, 76,
  \dodoi{10.3847/1538-3881/aa7de8}

\bibitem[{{Carry}(2012)}]{Car12}
{Carry}, B. 2012, \planss, 73, 98, \dodoi{10.1016/j.pss.2012.03.009}

\bibitem[{{Elkins-Tanton} {et~al.}(2014){Elkins-Tanton}, {Asphaug}, {Bell},
  {Bercovici}, {Bills}, {Binzel}, {Bottke}, {Jun}, {Marchi}, {Oh}, {Polanskey},
  {Weiss}, {Wenkert}, \& {Zuber}}]{Elk14}
{Elkins-Tanton}, L.~T., {Asphaug}, E., {Bell}, J., {et~al.} 2014, in Lunar and
  Planetary Science Conference, Vol.~45, Lunar and Planetary Science
  Conference, 1253

\bibitem[{{Farnocchia} {et~al.}(2015){Farnocchia}, {Chesley}, {Chamberlin}, \&
  {Tholen}}]{Far15}
{Farnocchia}, D., {Chesley}, S.~R., {Chamberlin}, A.~B., \& {Tholen}, D.~J.
  2015, \icarus, 245, 94, \dodoi{10.1016/j.icarus.2014.07.033}

\bibitem[{{Ferrais} {et~al.}(2020){Ferrais}, {Vernazza}, {Jorda}, {Rambaux},
  {Hanu{\v{s}}}, {Carry}, {Marchis}, {Marsset}, {Viikinkoski}, {Bro{\v{z}}},
  {Fetick}, {Drouard}, {Fusco}, {Birlan}, {Podlewska-Gaca}, {Jehin},
  {Bartczak}, {Berthier}, {Castillo-Rogez}, {Cipriani}, {Colas},
  {Dudzi{\'n}ski}, {Dumas}, {{\v{D}}urech}, {Kaasalainen}, {Kryszczynska},
  {Lamy}, {Le Coroller}, {Marciniak}, {Michalowski}, {Michel}, {Santana-Ros},
  {Tanga}, {Vachier}, {Vigan}, {Witasse}, \& {Yang}}]{Fer20}
{Ferrais}, M., {Vernazza}, P., {Jorda}, L., {et~al.} 2020, \aap, 638, L15,
  \dodoi{10.1051/0004-6361/202038100}

\bibitem[{{Fienga} {et~al.}(2003){Fienga}, {Bange}, {Bec-Borsenberger}, \&
  {Thuillot}}]{Fie03}
{Fienga}, A., {Bange}, J.~F., {Bec-Borsenberger}, A., \& {Thuillot}, W. 2003,
  \aap, 406, 751, \dodoi{10.1051/0004-6361:20030641}

\bibitem[{{Gal{\'a}d} \& {Gray}(2002)}]{Gal02}
{Gal{\'a}d}, A., \& {Gray}, B. 2002, \aap, 391, 1115,
  \dodoi{10.1051/0004-6361:20020810}

\bibitem[{{Granvik} {et~al.}(2009){Granvik}, {Virtanen}, {Oszkiewicz}, \&
  {Muinonen}}]{Gra09}
{Granvik}, M., {Virtanen}, J., {Oszkiewicz}, D., \& {Muinonen}, K. 2009,
  Meteoritics and Planetary Science, 44, 1853,
  \dodoi{10.1111/j.1945-5100.2009.tb01994.x}

\bibitem[{Johnson {et~al.}(2020)Johnson, Sori, \& Evans}]{Joh20}
Johnson, B.~C., Sori, M.~M., \& Evans, A.~J. 2020, Nature Astronomy, 4, 41,
  \dodoi{10.1038/s41550-019-0885-x}

\bibitem[{Lang {et~al.}(2010)Lang, Hogg, Mierle, Blanton, \& Roweis}]{Lan10}
Lang, D., Hogg, D.~W., Mierle, K., Blanton, M., \& Roweis, S. 2010, The
  Astronomical Journal, 139, 1782, \dodoi{10.1088/0004-6256/139/5/1782}

\bibitem[{{Russell} {et~al.}(2012){Russell}, {Raymond}, {Coradini}, {McSween},
  {Zuber}, {Nathues}, {De Sanctis}, {Jaumann}, {Konopliv}, {Preusker}, {Asmar},
  {Park}, {Gaskell}, {Keller}, {Mottola}, {Roatsch}, {Scully}, {Smith},
  {Tricarico}, {Toplis}, {Christensen}, {Feldman}, {Lawrence}, {McCoy},
  {Prettyman}, {Reedy}, {Sykes}, \& {Titus}}]{Rus12}
{Russell}, C.~T., {Raymond}, C.~A., {Coradini}, A., {et~al.} 2012, Science,
  336, 684, \dodoi{10.1126/science.1219381}

\bibitem[{{Russell} {et~al.}(2016){Russell}, {Raymond}, {Ammannito},
  {Buczkowski}, {De Sanctis}, {Hiesinger}, {Jaumann}, {Konopliv}, {McSween},
  {Nathues}, {Park}, {Pieters}, {Prettyman}, {McCord}, {McFadden}, {Mottola},
  {Zuber}, {Joy}, {Polanskey}, {Rayman}, {Castillo-Rogez}, {Chi}, {Combe},
  {Ermakov}, {Fu}, {Hoffmann}, {Jia}, {King}, {Lawrence}, {Li}, {Marchi},
  {Preusker}, {Roatsch}, {Ruesch}, {Schenk}, {Villarreal}, \&
  {Yamashita}}]{Rus16}
{Russell}, C.~T., {Raymond}, C.~A., {Ammannito}, E., {et~al.} 2016, Science,
  353, 1008, \dodoi{10.1126/science.aaf4219}

\bibitem[{{Shepard} {et~al.}(2017){Shepard}, {Richardson}, {Taylor},
  {Rodriguez-Ford}, {Conrad}, {de Pater}, {Adamkovics}, {de Kleer}, {Males},
  {Morzinski}, {Close}, {Kaasalainen}, {Viikinkoski}, {Timerson}, {Reddy},
  {Magri}, {Nolan}, {Howell}, {Benner}, {Giorgini}, {Warner}, \&
  {Harris}}]{She17}
{Shepard}, M.~K., {Richardson}, J., {Taylor}, P.~A., {et~al.} 2017, \icarus,
  281, 388, \dodoi{10.1016/j.icarus.2016.08.011}

\bibitem[{{Siltala} \& {Granvik}(2020)}]{Sil20}
{Siltala}, L., \& {Granvik}, M. 2020, \aap, 633, A46,
  \dodoi{10.1051/0004-6361/201935608}

\bibitem[{Vihola(2012)}]{Vih12}
Vihola, M. 2012, Statistics and Computing, 22, 997,
  \dodoi{10.1007/s11222-011-9269-5}

\bibitem[{{Viikinkoski} {et~al.}(2018){Viikinkoski}, {Vernazza}, {Hanu{\v s}},
  {Le Coroller}, {Tazhenova}, {Carry}, {Marsset}, {Drouard}, {Marchis},
  {Fetick}, {Fusco}, {{\v D}urech}, {Birlan}, {Berthier}, {Bartczak}, {Dumas},
  {Castillo-Rogez}, {Cipriani}, {Colas}, {Ferrais}, {Grice}, {Jehin}, {Jorda},
  {Kaasalainen}, {Kryszczynska}, {Lamy}, {Marciniak}, {Michalowski}, {Michel},
  {Pajuelo}, {Podlewska-Gaca}, {Santana-Ros}, {Tanga}, {Vachier}, {Vigan},
  {Warner}, {Witasse}, \& {Yang}}]{Vii18}
{Viikinkoski}, M., {Vernazza}, P., {Hanu{\v s}}, J., {et~al.} 2018, \aap, 619,
  L3, \dodoi{10.1051/0004-6361/201834091}

\end{thebibliography}
\bibliographystyle{aasjournal}

%% This command is needed to show the entire author+affiliation list when
%% the collaboration and author truncation commands are used.  It has to
%% go at the end of the manuscript.
%\allauthors

%% Include this line if you are using the \added, \replaced, \deleted
%% commands to see a summary list of all changes at the end of the article.
%\listofchanges

\end{document}